\documentclass{moriond}

\usepackage[T1]{fontenc}
\usepackage{xspace}

\newcommand{\gambit}{\textsf{GAMBIT}\xspace}

\newcommand{\GB}{\gambit}

\def\be{\begin{equation}}
\def\ee{\end{equation}}
\def\bea{\begin{eqnarray}}
\def\eea{\end{eqnarray}}

\newcommand{\Photo}{}

\begin{document}
\title{Global analyses of supersymmetry with GAMBIT}

\author{Pat Scott for the GAMBIT Collaboration}

\address{Fundamental Physics Section, Department of Physics, Imperial College London, Blackett Laboratory, Prince Consort Road, London SW7 2AZ, UK}

\maketitle\abstracts{
I provide a brief summary of the status of supersymmetric models with parameters defined at either the unification or weak scale, based on global fits using the \GB framework.
}

Supersymmetry (SUSY) realised at the weak scale remains one of the most popular scenarios for new physics, as it can solve many problems of the Standard Model (SM).  SUSY is constrained by many complementary probes, ranging from direct searches at the LHC and LEP to flavour physics, precision measurements, direct and indirect searches for the lightest neutralino as dark matter (DM), and its thermal relic abundance.

We have analysed a number of SUSY scenarios \cite{CMSSM,MSSM} using the new global-fitting framework \GB \cite{gambit,darkbit,colliderbit,flavbit,SDPBit,scannerbit,SSDM}, producing the most advanced and comprehensive assessment of SUSY to date. The Minimal Supersymmetric SM (MSSM) may be described in terms of Lagrangian parameters defined directly at the weak scale, or at some other higher scale, where it is typically postulated that some of the parameters unify.  We carried out global fits of the Constrained MSSM (CMSSM) and the first and second non-universal Higgs mass (NUHM1/2) models, which respectively have 4, 5, and 6 parameters defined at the scale of grand unification \cite{CMSSM}.  We also carried out a 7-parameter scan of a version of the MSSM with the parameters defined at the weak scale \cite{MSSM}.

In each case, we also varied 5 nuisance parameters in the fits: the strong coupling $\alpha_s$, the top mass $m_t$, two nuclear matrix elements $\sigma_s$ and $\sigma_l$ (relevant for direct detection of DM), and the local density of DM $\rho_0$.  For these fits, we employed the differential evolution sampler \textsf{Diver} \cite{scannerbit} to efficiently and accurately map the likelihood function.  The full set of likelihood functions that we included in these fits can be found in the relevant publications \cite{CMSSM,MSSM} (Tables 5 and 3, respectively). One particular point of note is that we treat the observed abundance of DM \cite{Planck15cosmo} as an upper limit on the relic density of neutralinos, rather than demanding that they constitute the entirety of DM.

\begin{figure}[tbp]
\centering
    \includegraphics[width=0.46\textwidth]{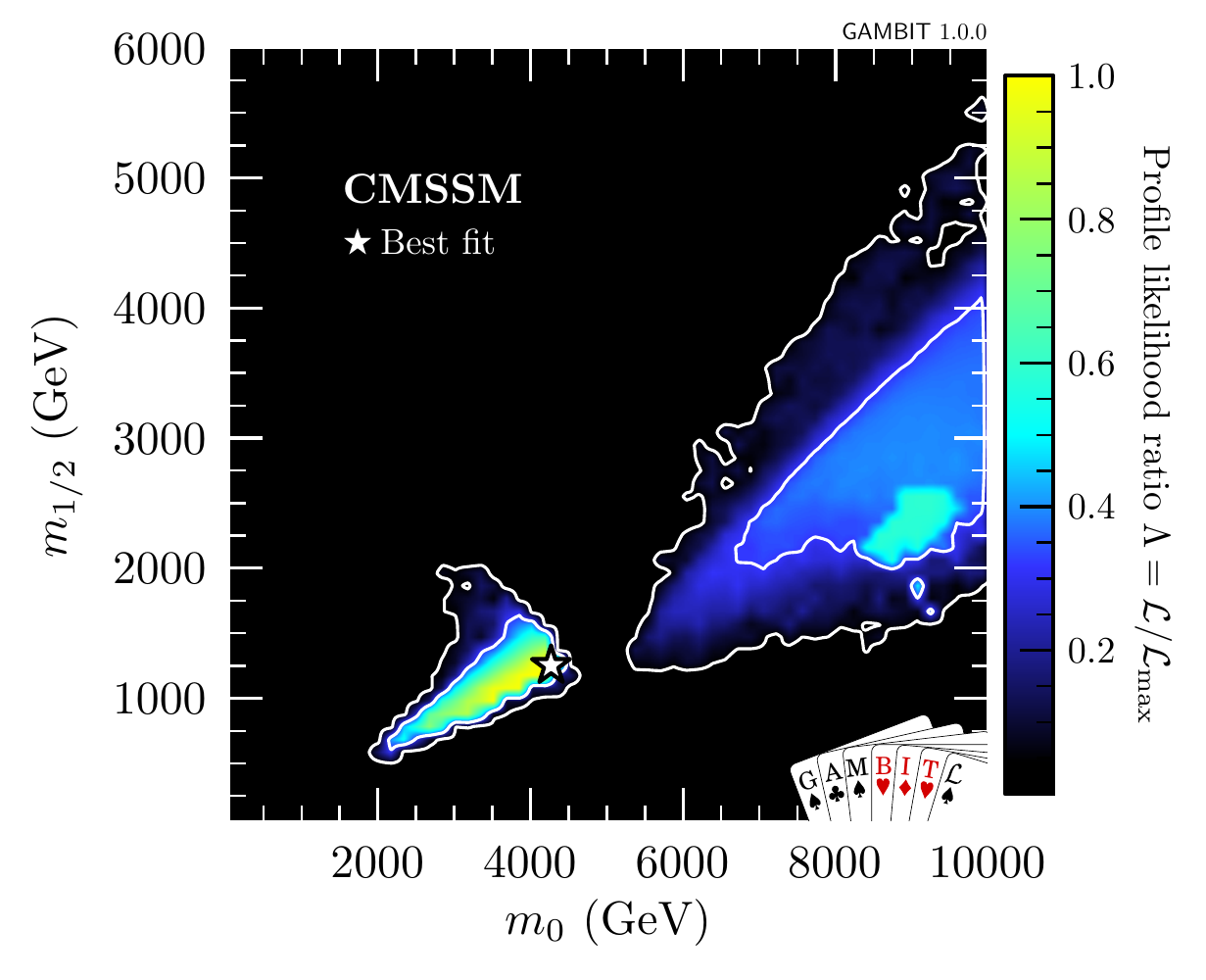}\hspace{3mm}%
    \includegraphics[width=0.46\textwidth]{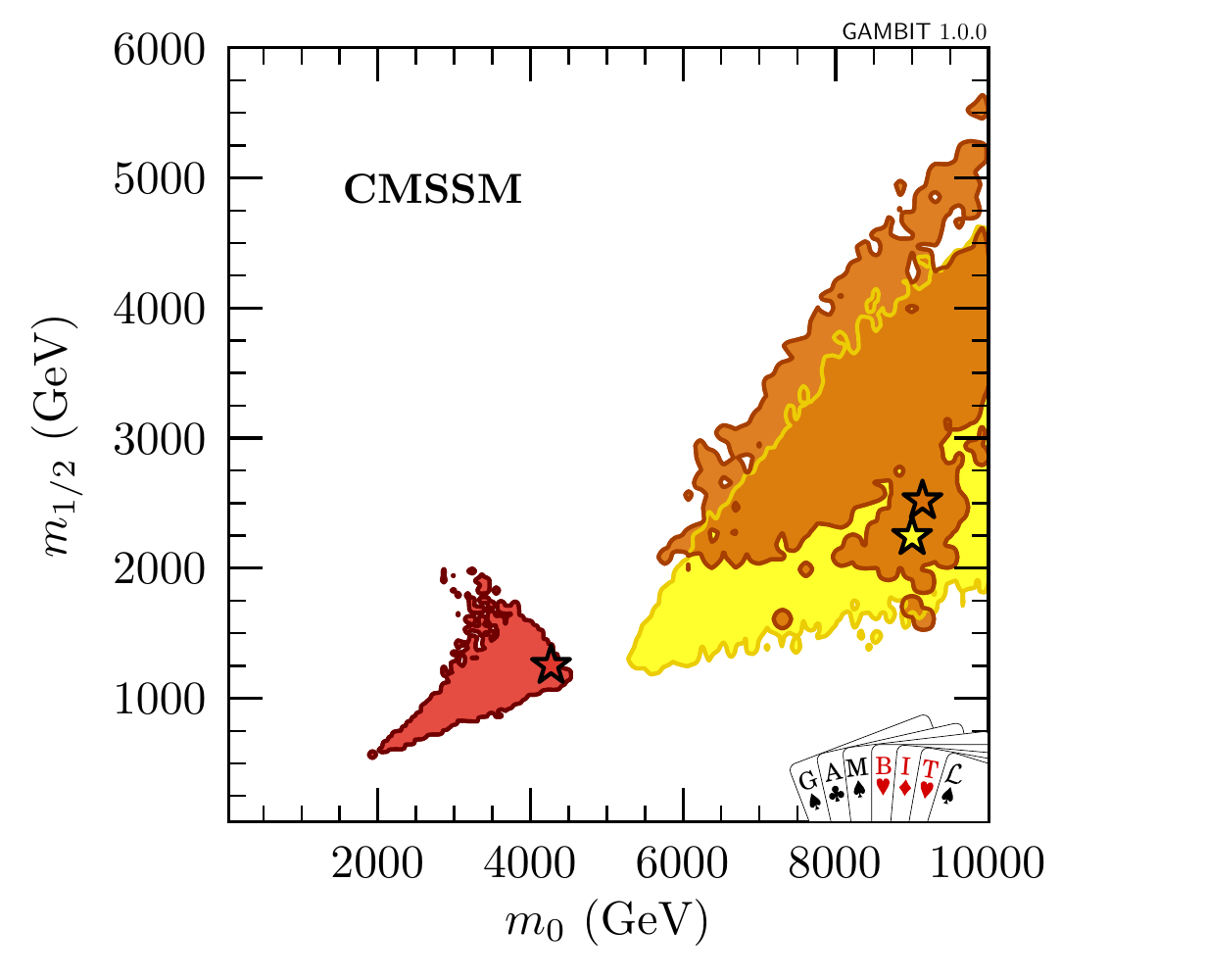}\\
    \includegraphics[width=0.46\textwidth]{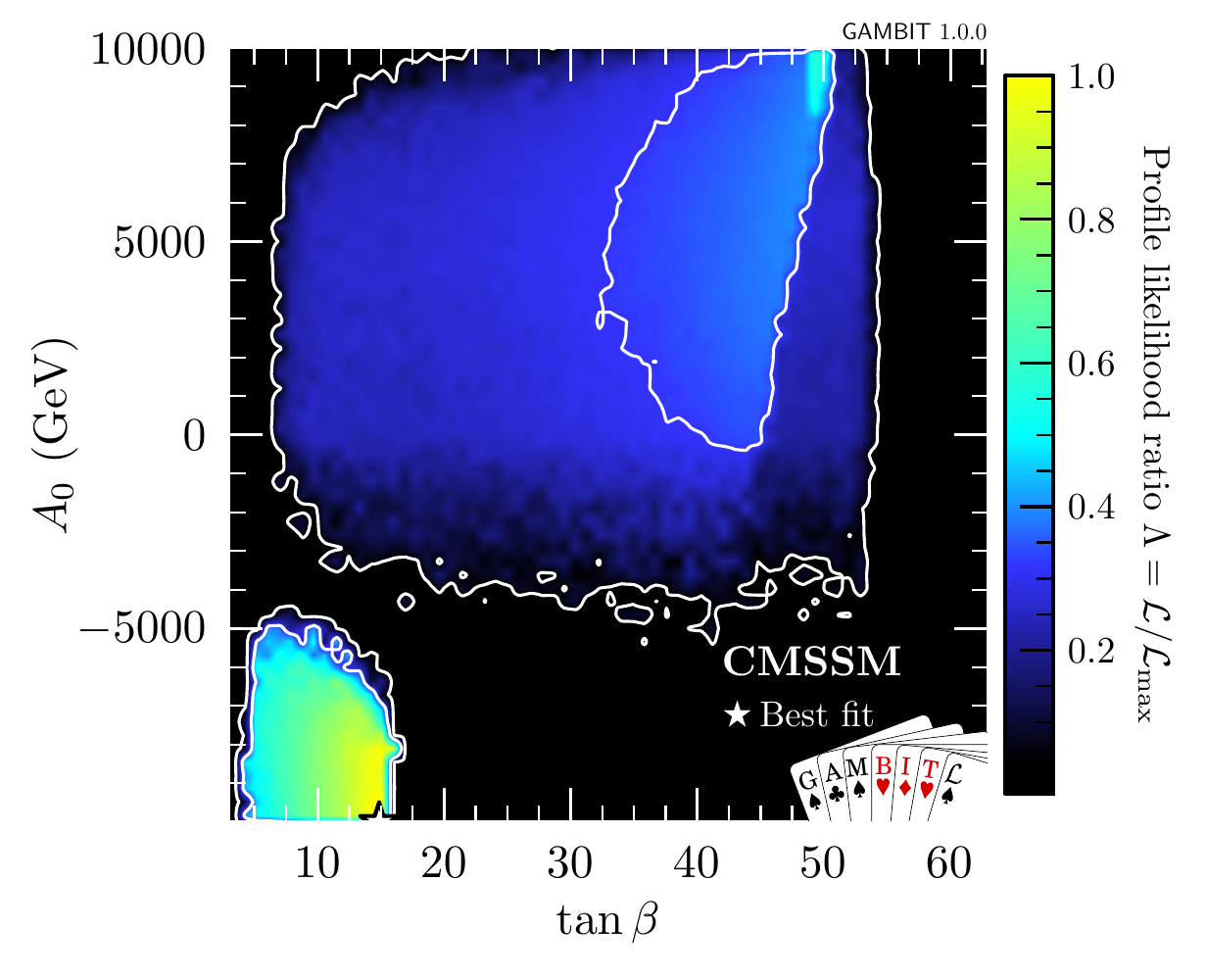}\hspace{3mm}%
    \includegraphics[width=0.46\textwidth]{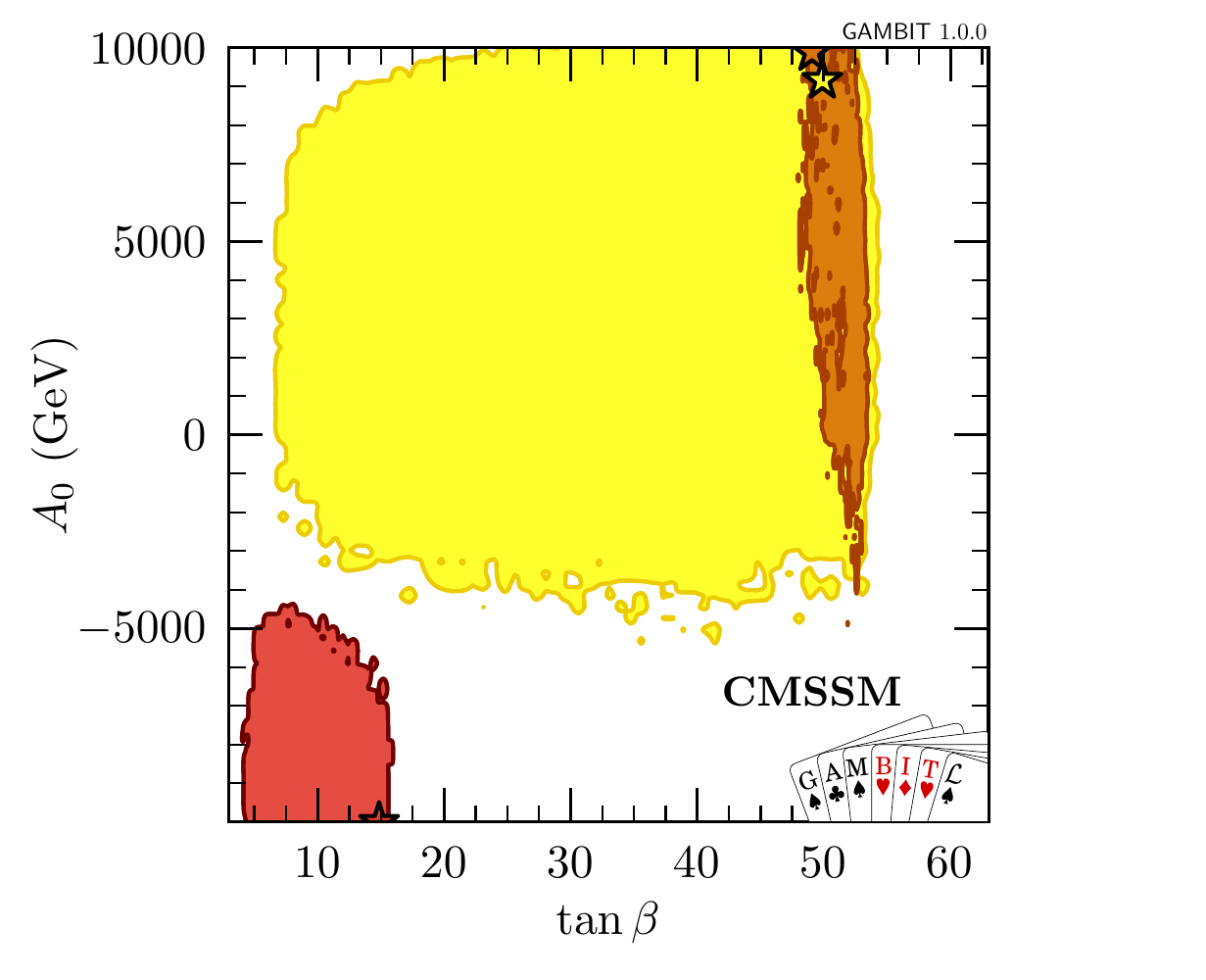}\\
    \includegraphics[width=0.46\textwidth]{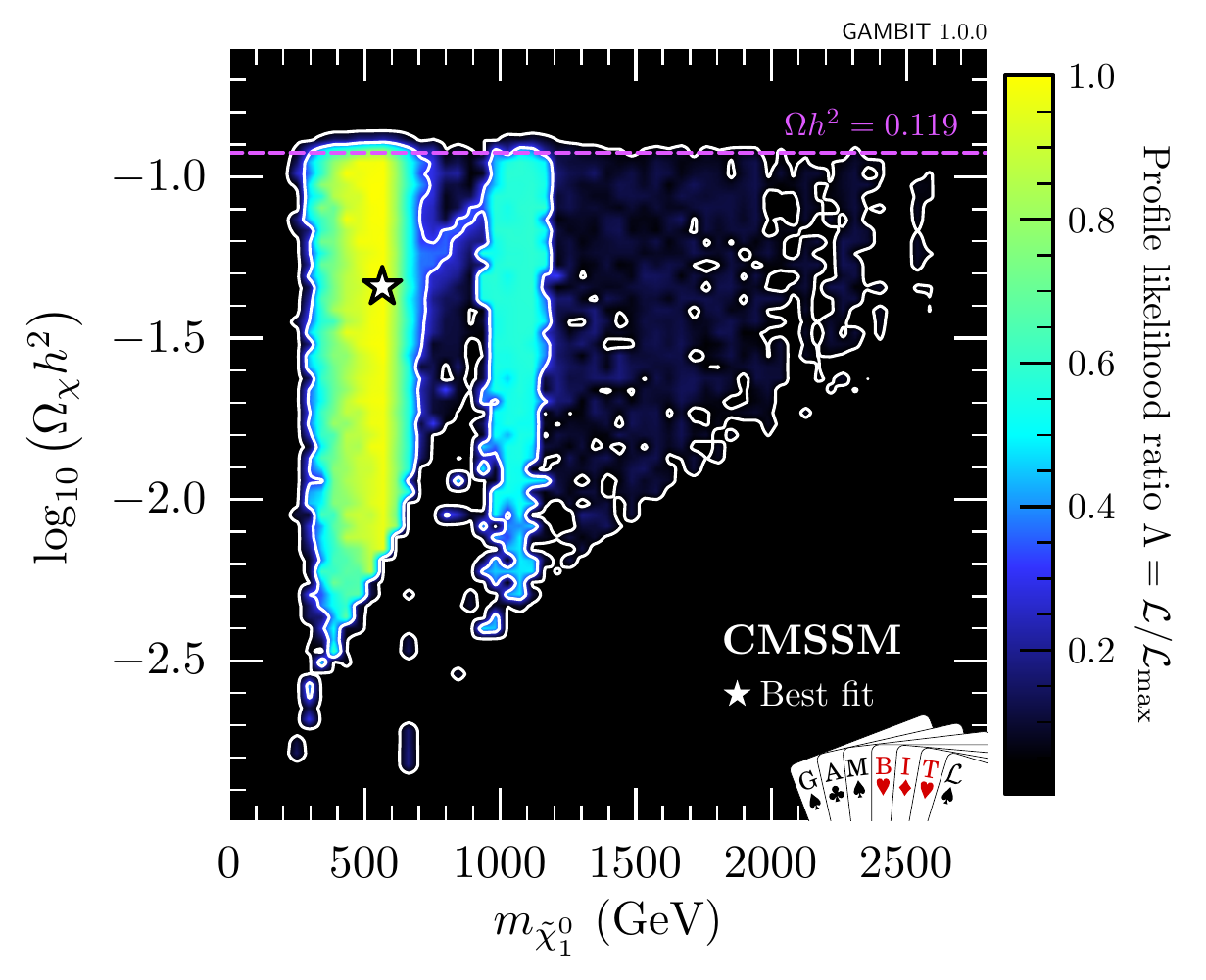}\hspace{3mm}%
    \includegraphics[width=0.46\textwidth]{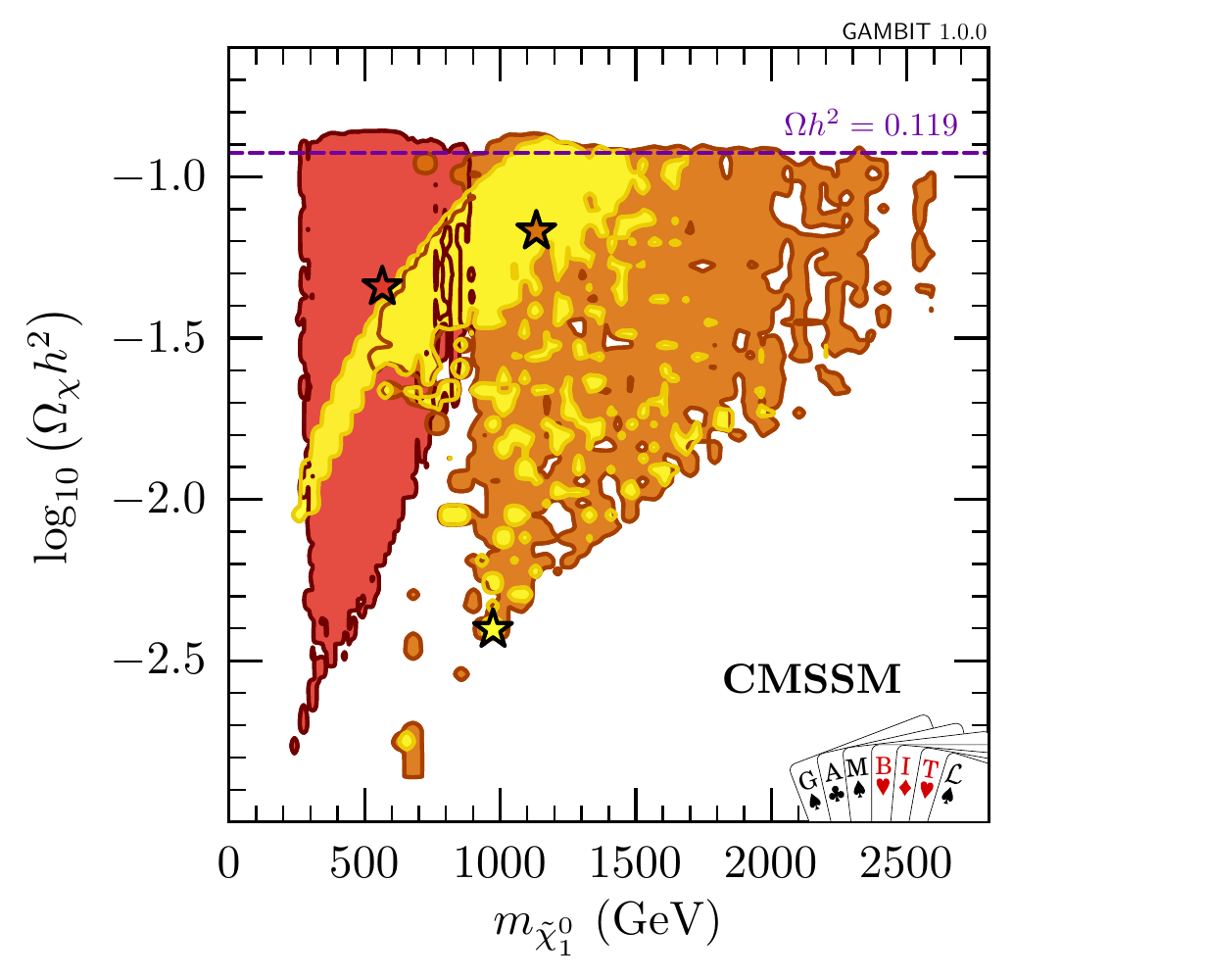}\\
    \includegraphics[width=0.46\textwidth]{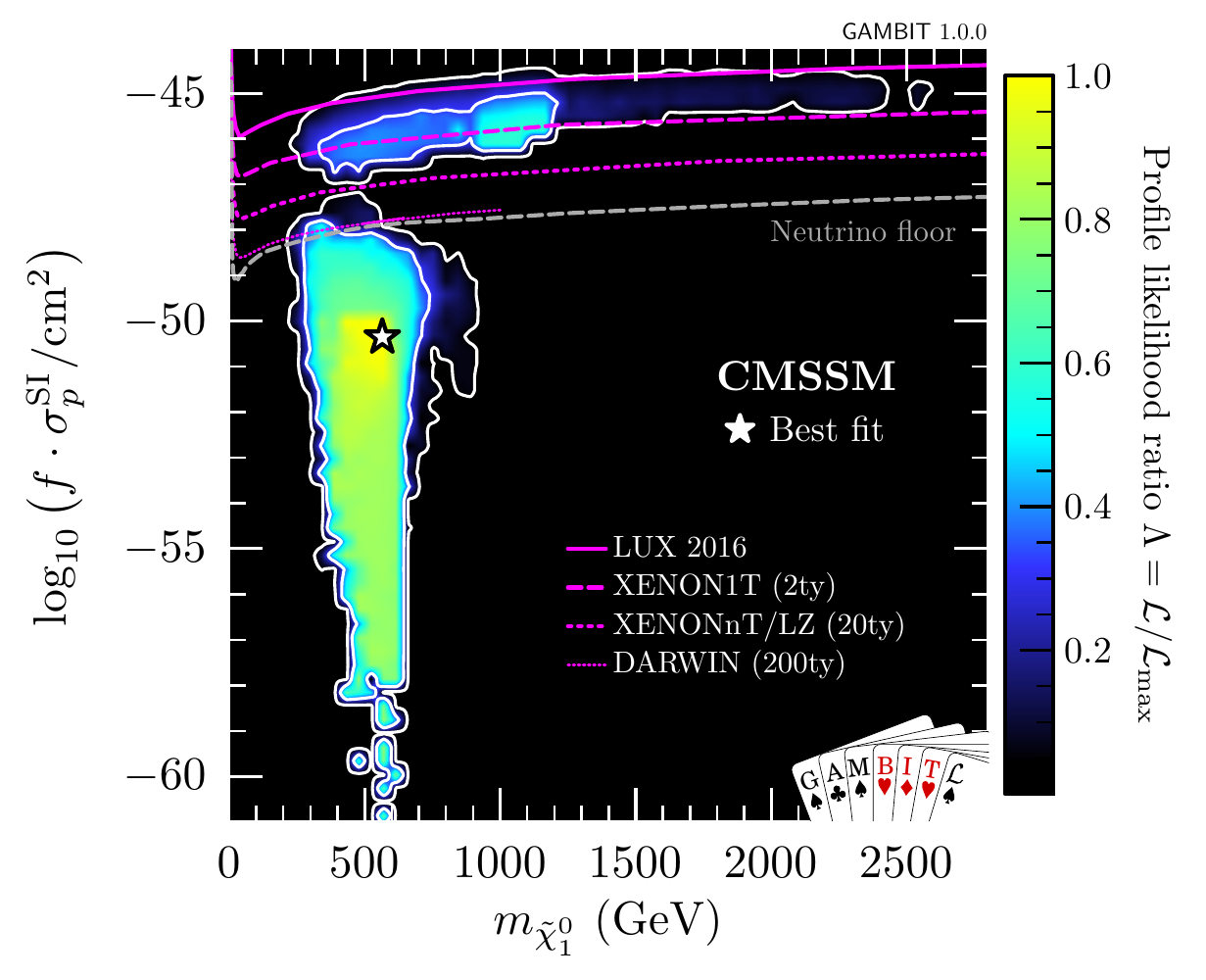}\hspace{3mm}%
    \includegraphics[width=0.46\textwidth]{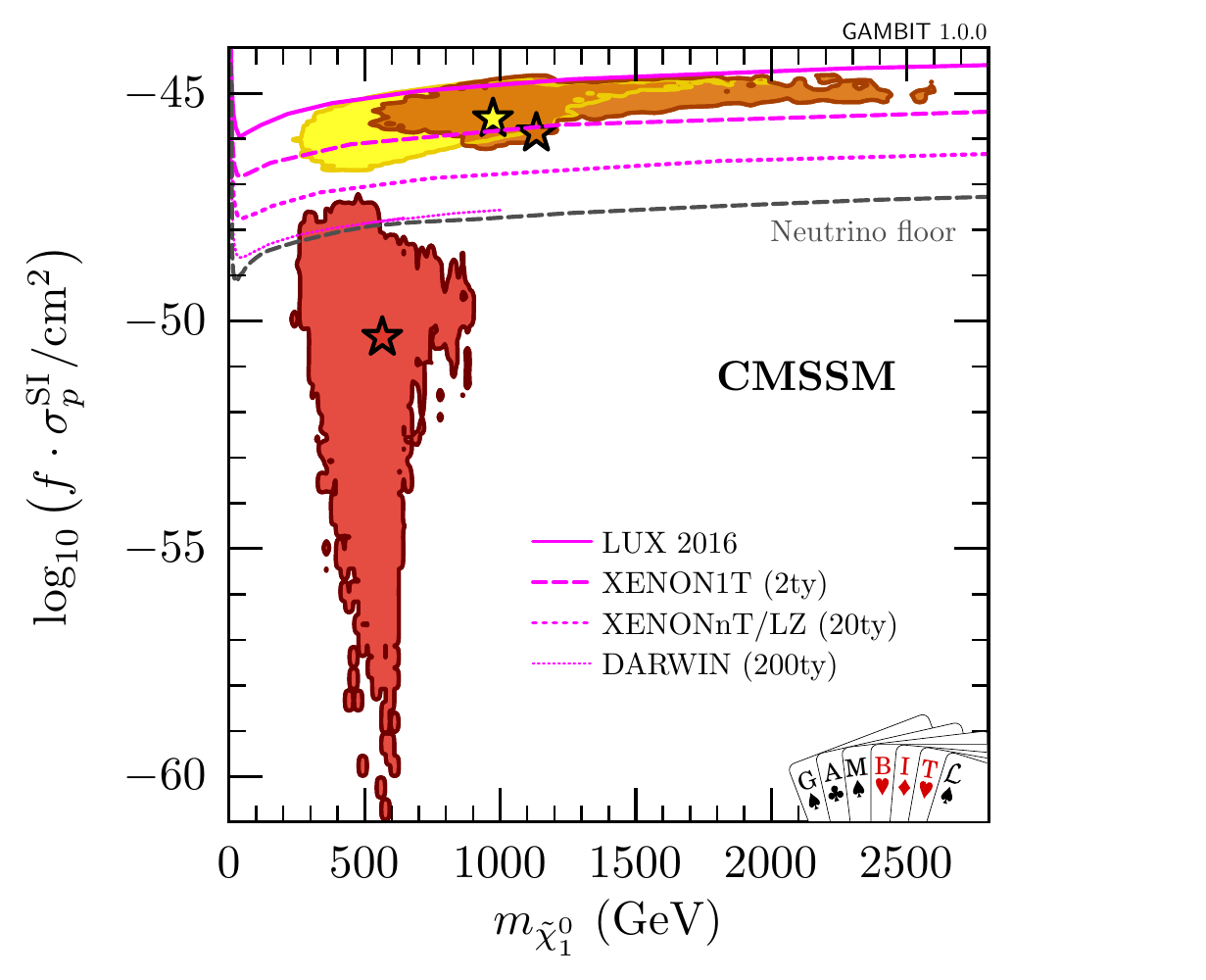}\\
    \includegraphics[height=0.3cm]{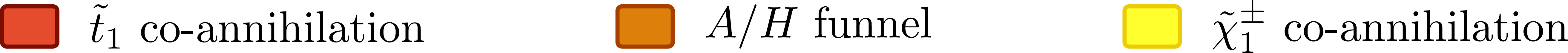}
\caption{Global fits of the CMSSM \protect\cite{CMSSM}, showing pairs of fundamental parameters, the relic density, the lightest neutralino mass and the spin-independent nuclear scattering cross-section, rescaled for the fraction $f$ of DM made up by neutralinos.  Left panels show profile likelihoods; white contours correspond to 1 and 2$\sigma$ confidence regions. Right panels show 2$\sigma$ regions colour-coded according to the most active mechanisms involved in determining the relic density of the lightest neutralino.}
\label{fig1}
\end{figure}

\begin{figure}[tbp]
\centering
    \includegraphics[width=0.46\textwidth]{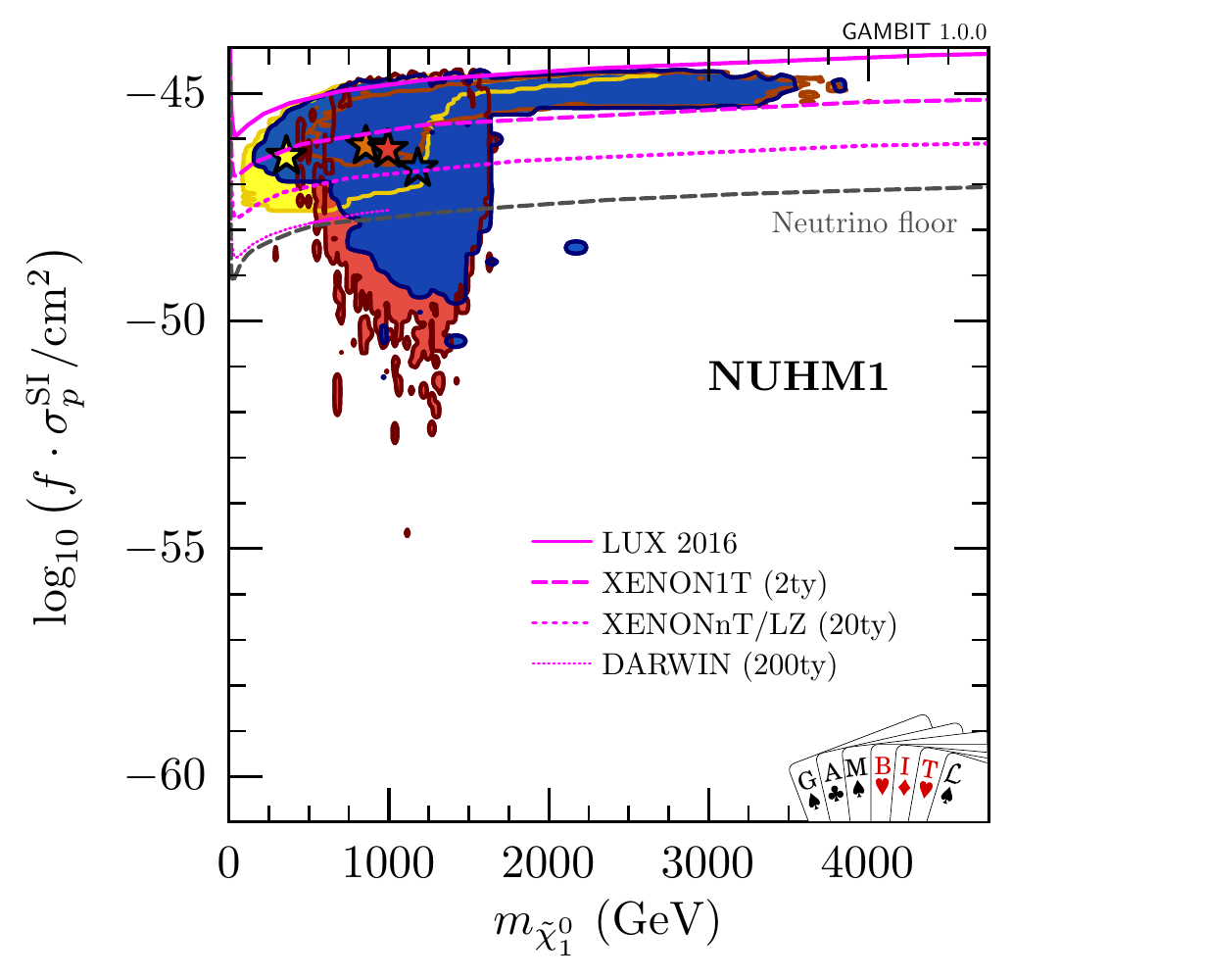}\hspace{3mm}%
    \includegraphics[width=0.46\textwidth]{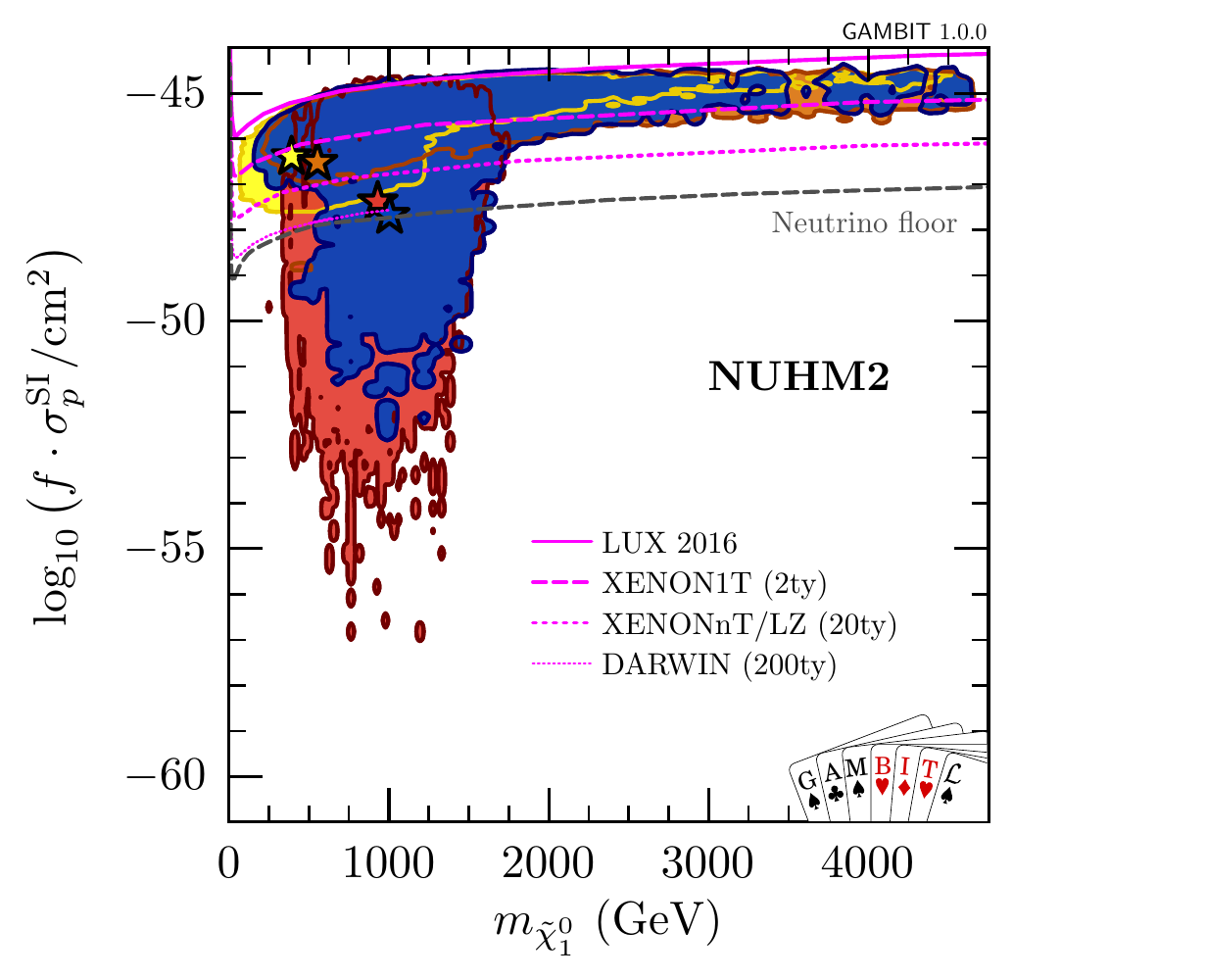}\\
    \centering
    \includegraphics[height=0.3cm]{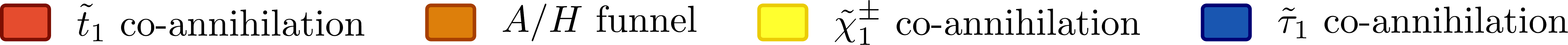}
\caption{Global fits of the NUHM1 and NUHM2 models \protect\cite{CMSSM}, showing spin-independent nuclear scattering cross-sections, rescaled for the fraction $f$ of DM made up by neutralinos.}
\label{fig2}
\end{figure}

Results of the CMSSM fits can be seen in Fig.\ \ref{fig1}. Here we see three distinct solutions, corresponding to three different mechanisms for obtaining an acceptable relic density: Higgsino DM (with chargino co-annihilation), stop co-annihilation, and annihilation through a heavy Higgs resonance.  The final row shows current and projected limits on spin-independent DM-nucleon scattering from direct detection.  Here we show rescaled spin-independent scattering cross-sections, which include a factor to account for the suppression of the local DM density due to the fraction $f = \Omega_\chi / \Omega_\mathrm{DM}$ of DM in neutralinos.  These panels can be compared with Fig.\ \ref{fig2}, which shows the same quantity in the NUHM1 and NUHM2 models, illustrating the presence of an additional stau co-annihilation solution in the latter two cases, leading to a large volume of the parameter space with extremely small scattering cross-sections.

\begin{figure}[tbp]
\centering
    \includegraphics[width=0.46\textwidth]{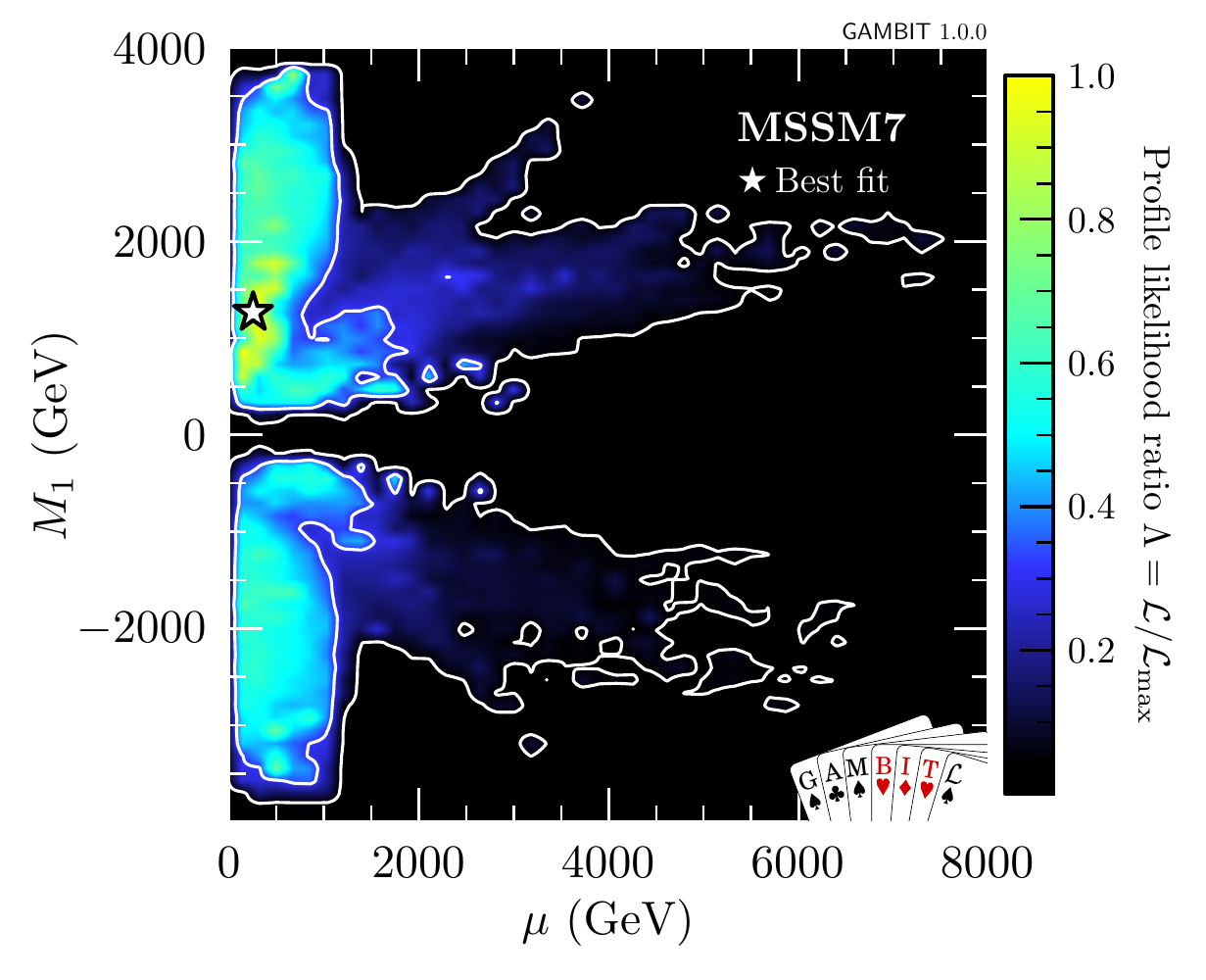}\hspace{3mm}%
    \includegraphics[width=0.46\textwidth]{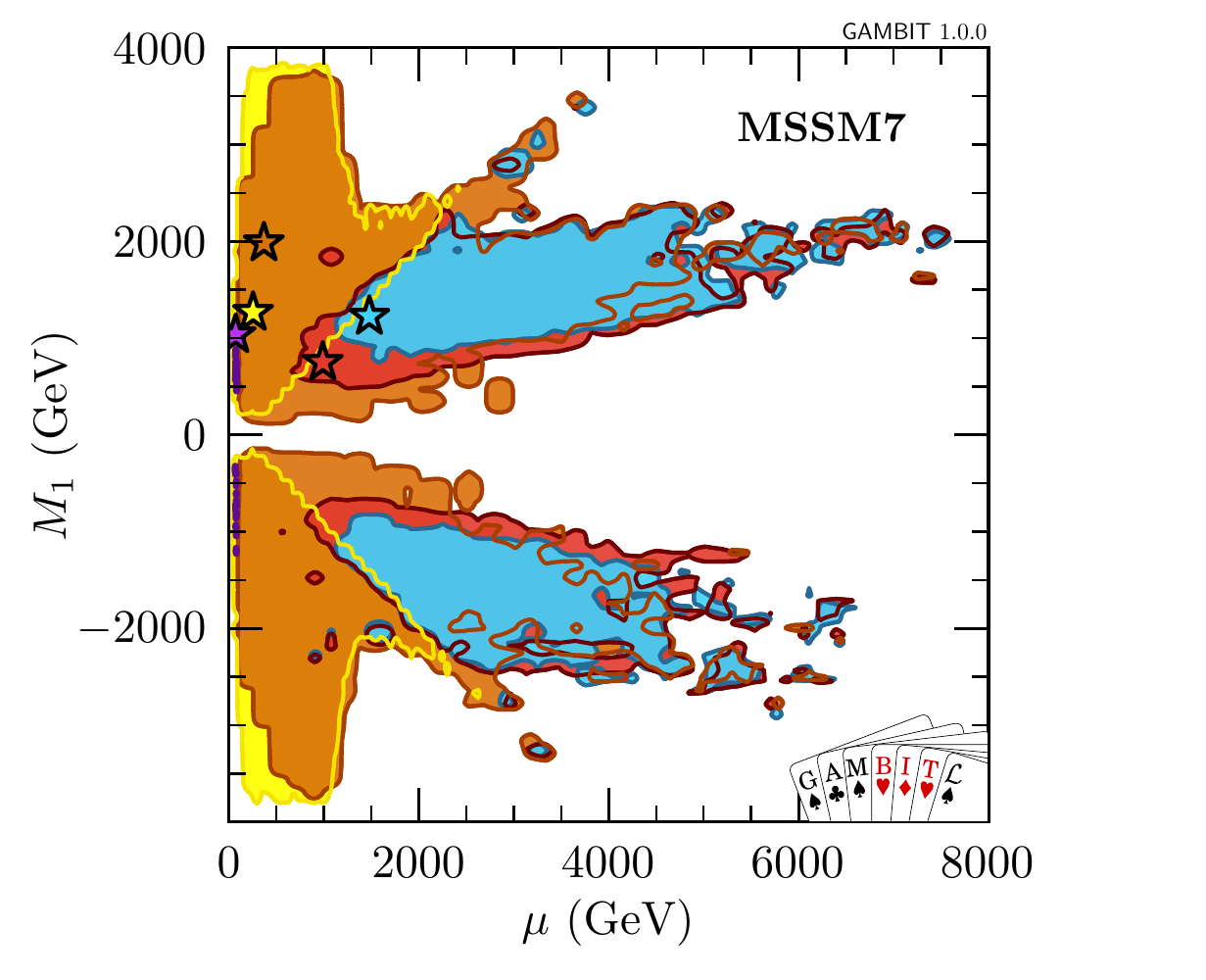}\\
    \includegraphics[width=0.46\textwidth]{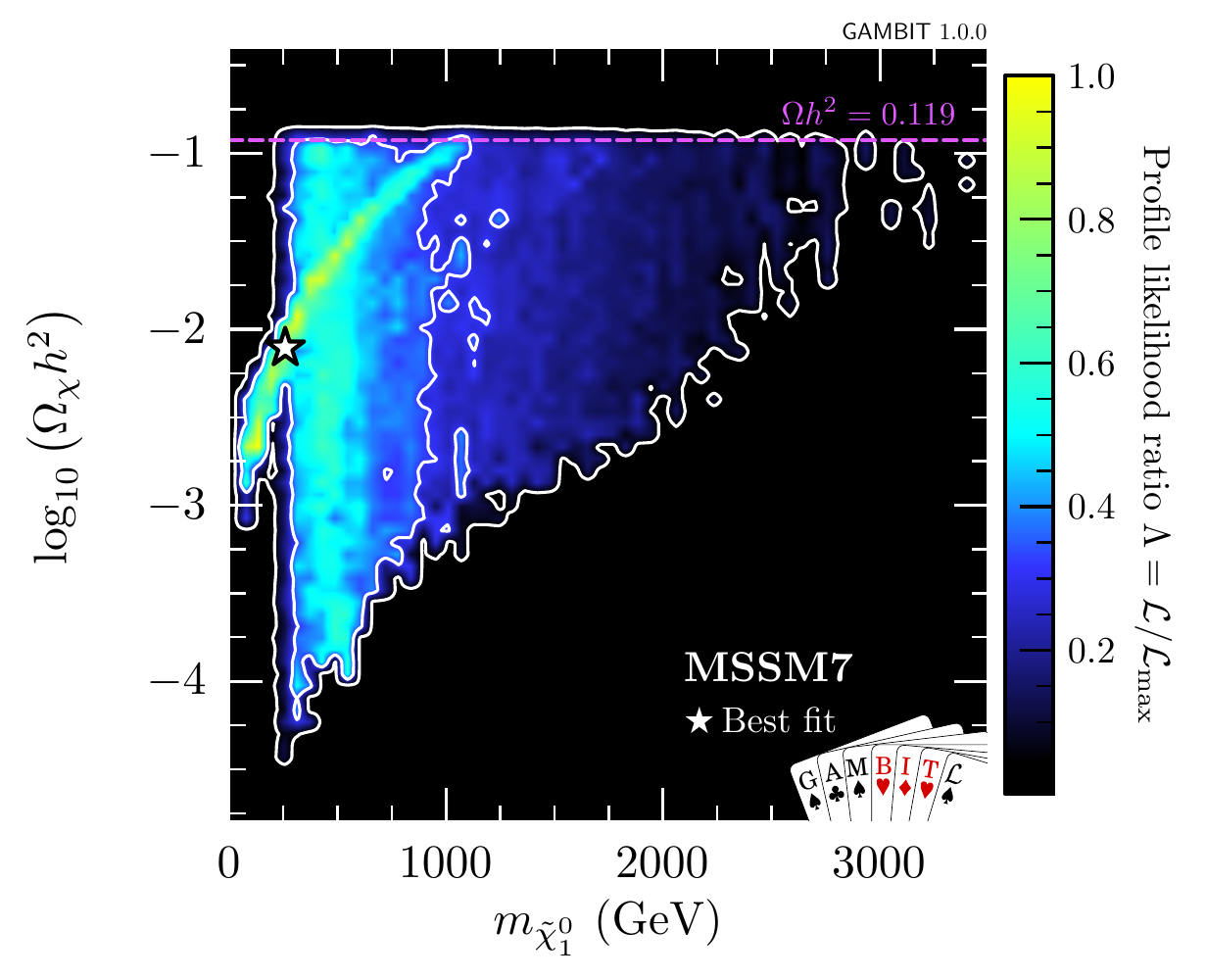}\hspace{3mm}%
    \includegraphics[width=0.46\textwidth]{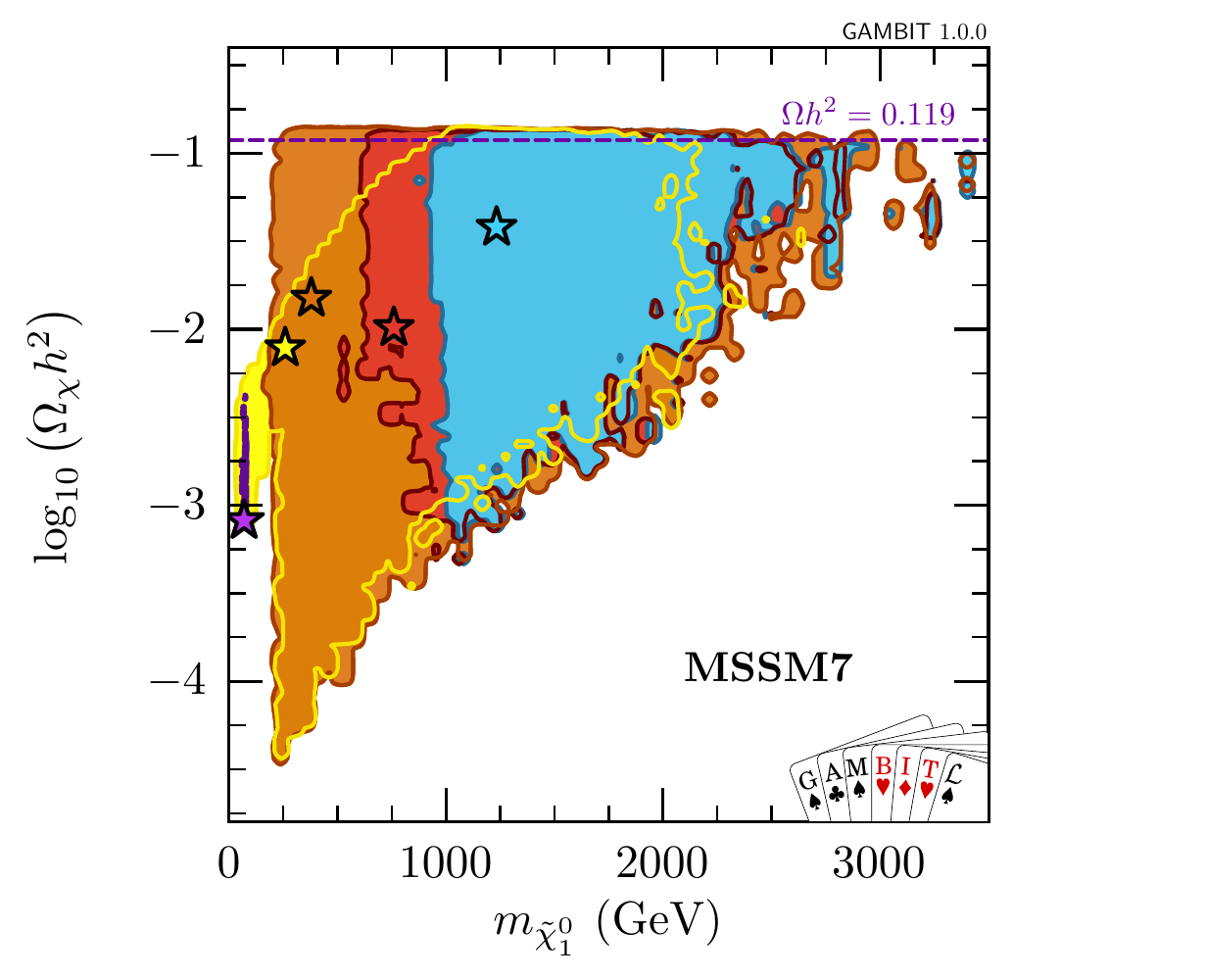}\\
    \includegraphics[width=0.46\textwidth]{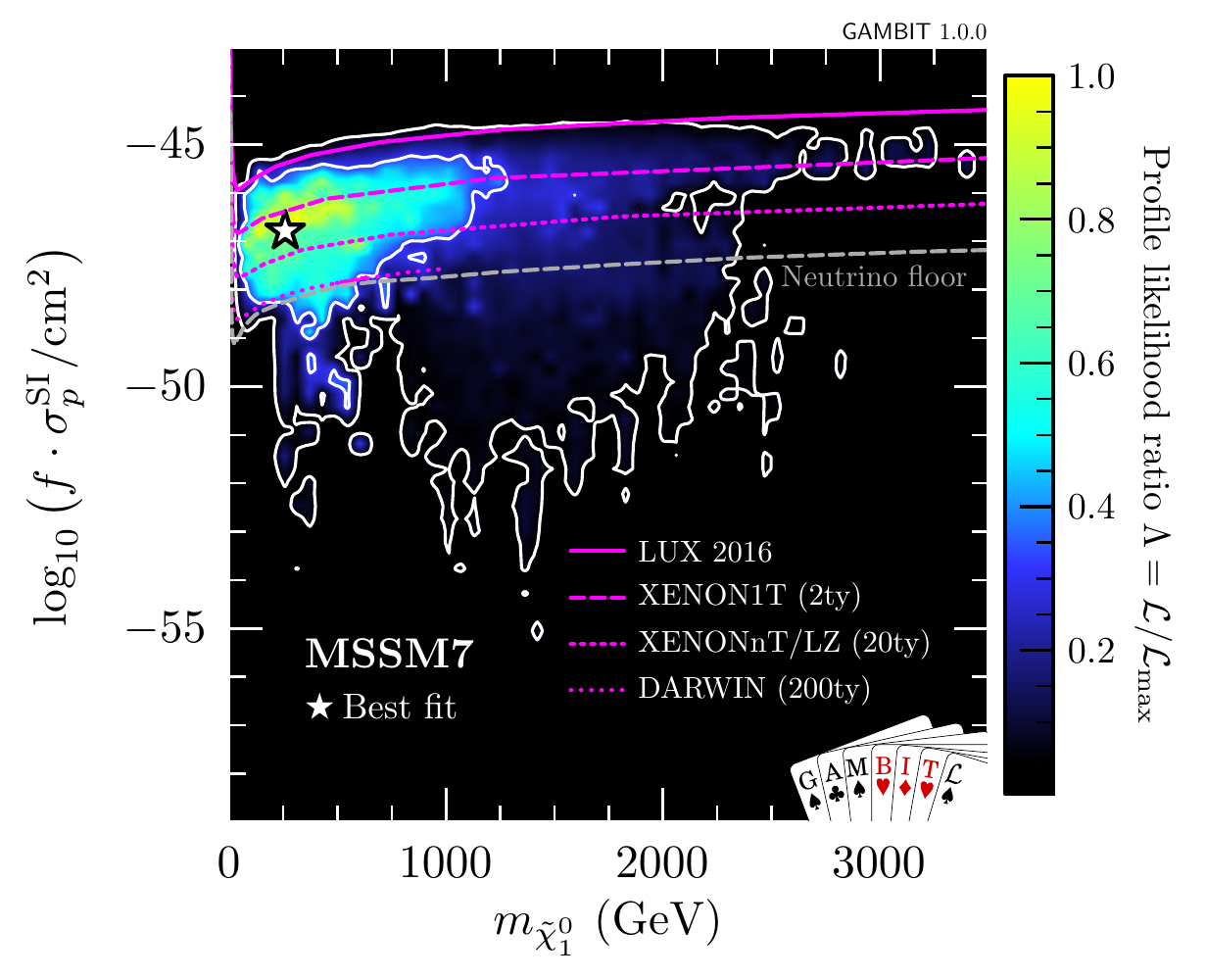}\hspace{3mm}%
    \includegraphics[width=0.46\textwidth]{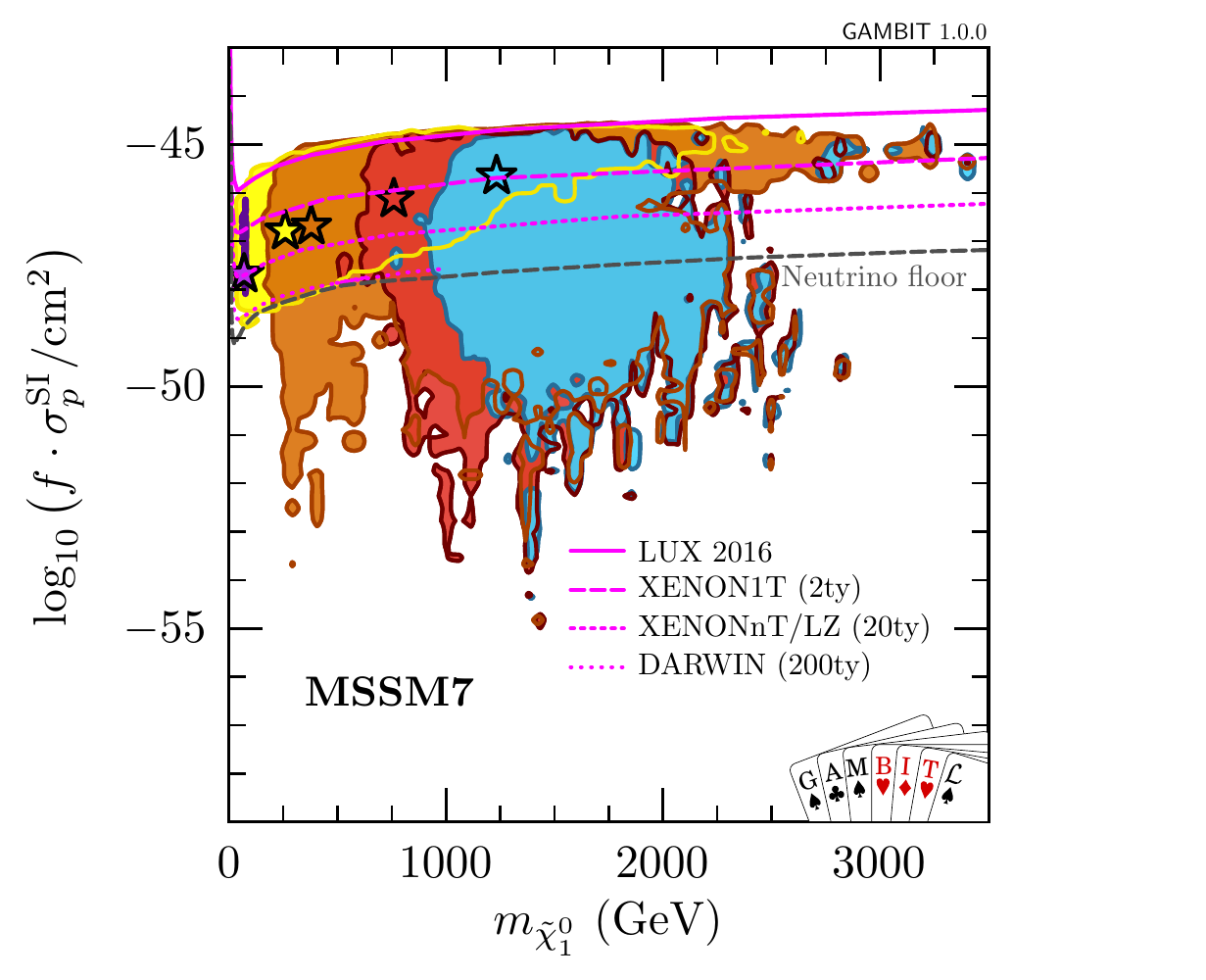}\\
    \includegraphics[width=0.46\textwidth]{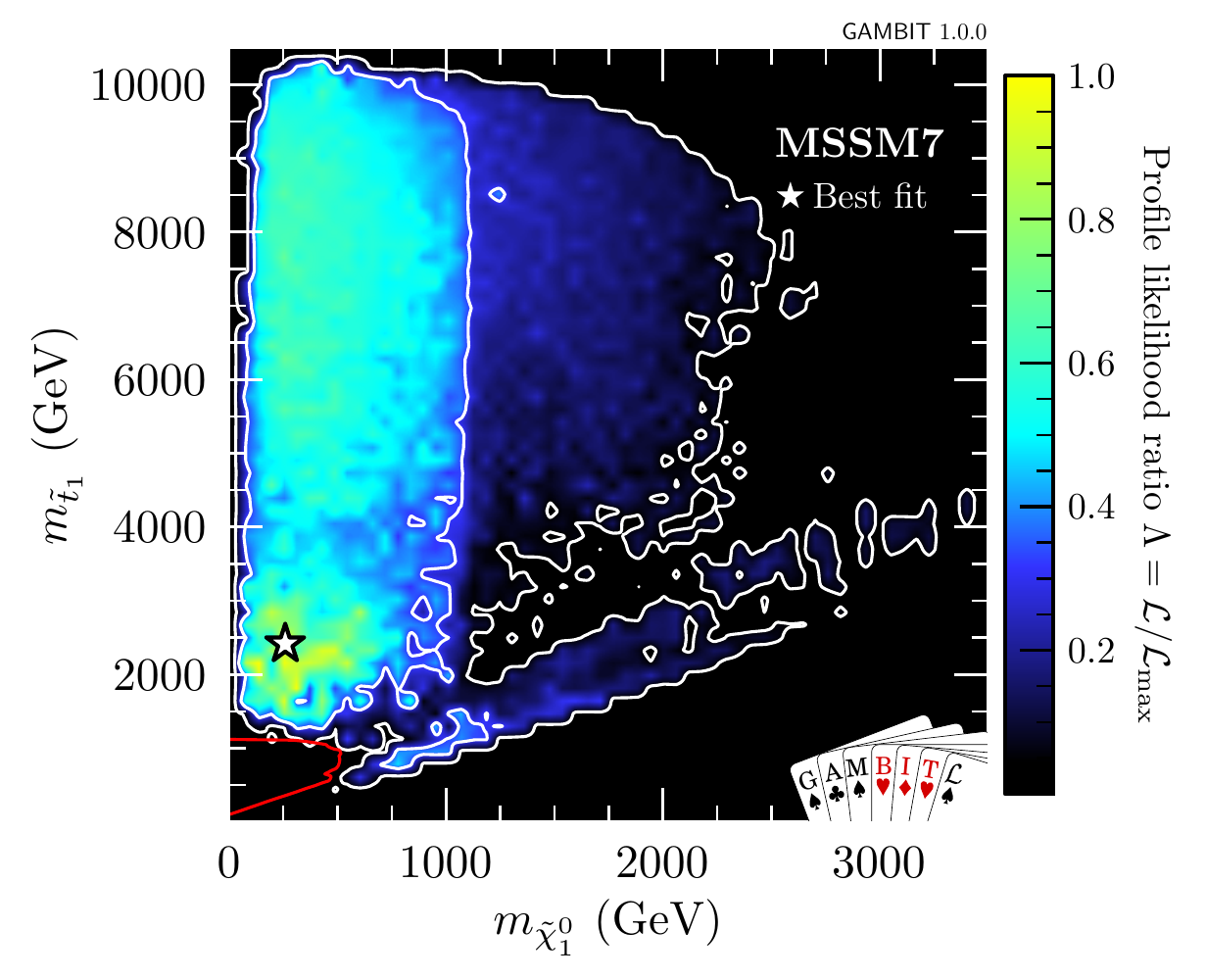}\hspace{3mm}%
    \includegraphics[width=0.46\textwidth]{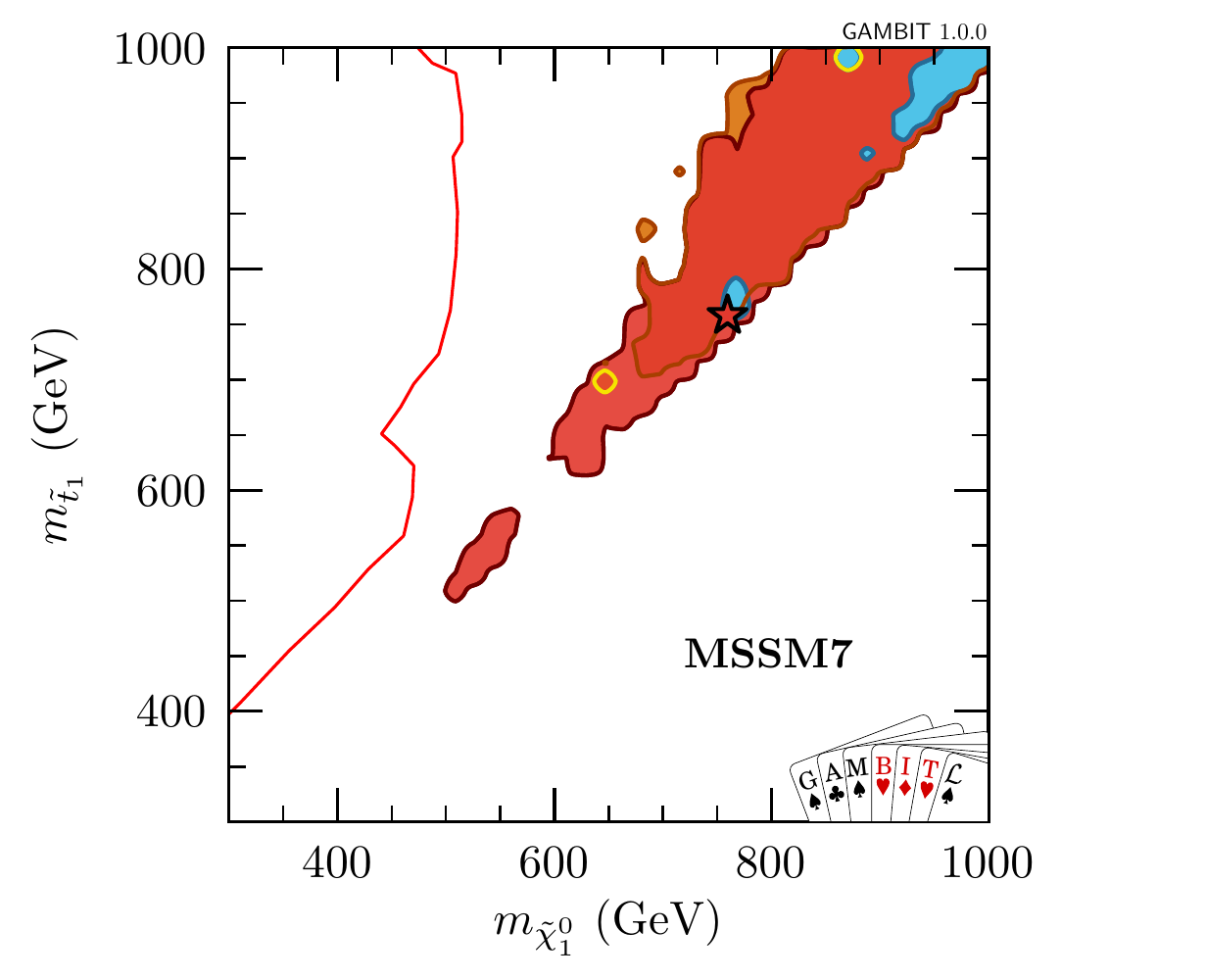}\\
    \includegraphics[height=0.3cm]{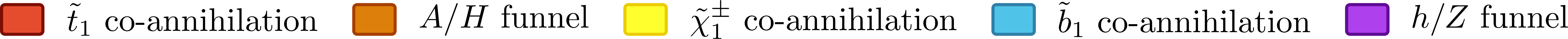}
\caption{Global fits of the MSSM-7 \protect\cite{MSSM}, showing fundamental parameters, the relic density, the spin-independent nuclear scattering cross-section (rescaled for the fraction $f$ of DM made up by neutralinos), and stop and neutralino masses.  Left and right panels as per Fig.\ \protect\ref{fig1}.  Bottom panels include a red line corresponding to the sensitivity of the CMS Run II simplified model limit from stop pair production \protect\cite{CMSStopSummary}.}
\label{fig3}
\end{figure}

The status of the 7 parameter weak-scale MSSM is summarised in Fig.\ \ref{fig3}, where 5 different mechanisms can be distinguished: stop and sbottom co-annihilation, Higgsino DM (chargino co-annihilation), and both light and heavy Higgs resonances.  The lowermost panels compare the preferred stop co-annihilation region to the current reach of the CMS Run II simplified model limit from stop pair production \cite{CMSStopSummary}.  Although such compressed spectra are going to prove mostly impossible to probe with direct and indirect detection, this figure indicates that the LHC is within striking reach of some of the most interesting solutions in this model, and may be able to probe this region in Run III.

\section*{Acknowledgments}

I am supported by STFC (ST/K00414X/1, ST/P000762/1, ST/L00044X/1), and thank my collaborators in GAMBIT for their contributions to the results discussed here.

\vspace{-2mm}
\section*{References}

\bibliography{R1.5}

\end{document}